# Design And Modelling An Attack on Multiplexer Based Physical Unclonable Function

Abhijith Manchikanti Venkata [1], Dinesh Reddy Jeeru [2], Vittal K.P.[3]

[1] *PG Scholar, Department of E.C.E, B.M.S College of Engineering, Bangalore, India,*
[2] *Assistant Professor, Department of E.C.E, B.M.S College of Engineering, Bangalore, India,*
[3] *Professor, Department of E.E.E, National Institute of Technology Karnataka, Surathkal, India.*

**Abstract**

*This paper deals with study of the physical unclonable functions and specifically the design of arbiter based PUF (APUF) and extends the work on different types of attacks on the PUF designs to break the security of the device, which includes advanced computational algorithms. Machine learning (ML) based attacks are successful in attacking existing designs. So in this, the resistance of the modified, proposed design of APUF is examined by modelling the attack based on the logistic regression a ML-based algorithm. The design is validated on Basys-3 Artix -7 FPGA board with a part number (xc7a35tcpg236-1).*

## I. INTRODUCTION

Nowadays, electronic devices are ruling the world mostly in every domain. Several applications are introduced in each field; the devices may be IoT devices, swiping cards, smart devices, and includes defense and medical applications. As the innovation increases at an equal rate security threat also rise. So security becomes the most important concern in the present digital world. Typically, for securing the secret key of the particular device, the key will be stored in device memory itself. But due to the advancement of several computational methods along with traditional methods makes the attackers quite successful in tracing the data which is stored in memory and data becomes public. To overcome that, a new security primitive physical unclonable function has come into existence. In this method, the generation of secret keys happens with the help of process variations of the integrated circuit. As it depends on the variations in the fabrication process, from device to device the generated data will be unique. So there is no possibility of cloning of data. As the name suggests that the function is not clonable on the physical system.

## II. RELATED WORK

The physical unclonable function is the advanced hardware-dependent security primitive for electronic devices. To provide security, different types of physical unclonable function design structures have come into existence. These are classified based on their sources of randomness generation such as intrinsic and extrinsic. Based on the challenges or inputs applied to design, the designs are classified into strong PUF and weak PUF. If there are more number of challenges applied then that design is identified as strong PUF whereas if there are less in number like one challenge then the design is weak PUF. As discussed earlier, the intrinsic type PUFs are memory-based designs like SRAM PUF, DRAM PUF, and delay based designs such as Arbiter PUF, RO PUF, and extrinsic are such as optical and coating PUF designs [4]. Apart from the security these deigns have an application of device identification [9].

To break the security designed using PUFs, several methods are introduced which include invasive, side channel, linear delay model; wear out based attacks, and different types of algorithms [1], [7], and [8]. Other attacks are based on advanced computational algorithms which include logistic regression (LR), support vector machines (SVM), and artificial neural network based attacks [5]. The generation of random challenges helps to resist the attacks [6].

In this paper, the following sections deal with Methodology which describes the design principle of the arbiter based PUF and proposed design of PUF. In the further sections attack modeling on the proposed PUF and the experimental results are shown.

## III. METHODOLOGY

### A. Arbiter based Physical Unclonable Function:

A function which takes n-inputs as challenges (C) and provides a single output as the response(R). It was the first design proposed under the silicon-based design PUF's. The fundamental idea of this PUF design is to generate the race conditions between two path signals; the input is provided as a challenge and generates a single bit response output. The design works on the delay variations across the chip. During chip fabrication, variations in manufacturing will have an impact on the physical parameters, which helps to determine the exact delay of each path and causes a small random offset between the two delays. This PUF circuit contains 'n' consecutive multiplexer blocks and an arbiter element which is typically a flip flop. Each multiplexer block contains two multiplexers of 2-input with a common select bit for both. The inputs are provided to these selected bits. Based on the logic value at the select line, the signal is travel to that path.

As seen from fig (1), there are 'n' multiplexer switch blocks where each block outputs are acts as inputs to the further connected block. At the final stage, the outputs act as inputs to the arbiter element. For the initial stage block, the inputs are tied together and provided with an enable signal as input. So the pulse signal is traveling through the two paths. Typically, the two paths have to be similar but due to fabrication variations little amount difference will be introduced and there will be a delay in reaching the last element. At last, the arbiter element D-flip flop decides the output bit as '0' or '1' based on the path arrivals at the input.

*B. Proposed Design:*

As previously mentioned, the conventional arbiter PUF design is secure and it has the generation possibility of an exponential number of challenge-response pairs, so the prediction of the design challenges is difficult for attackers. But due to the advancement of several computational algorithms, there is a possibility of breaking the design with a good prediction rate. As per [2] the attacks for designs are modeled with the support of computational algorithms, the practical results are extracted with a prediction rate of challenges in high percentage for so many existing PUF designs, which include the conventional Arbiter based PUF design.

In this case, to improve the resistance of design some modifications are needed to existed design which makes the design strong and makes it difficult for the attackers to predict the model.

Here, modification is to generate an equal number of responses to the number of challenges applied. The design is modified with N-bit challenges to produce an N-bit response instead of a single response. The design functionality was written in Verilog language and implemented with the help of placement constraints and macro in the Xilinx vivado design suite to get the required results [3] as shown in fig (6).

The Proposed design is shown in fig (2). The proposed design experimental results will be discussed in a further section.

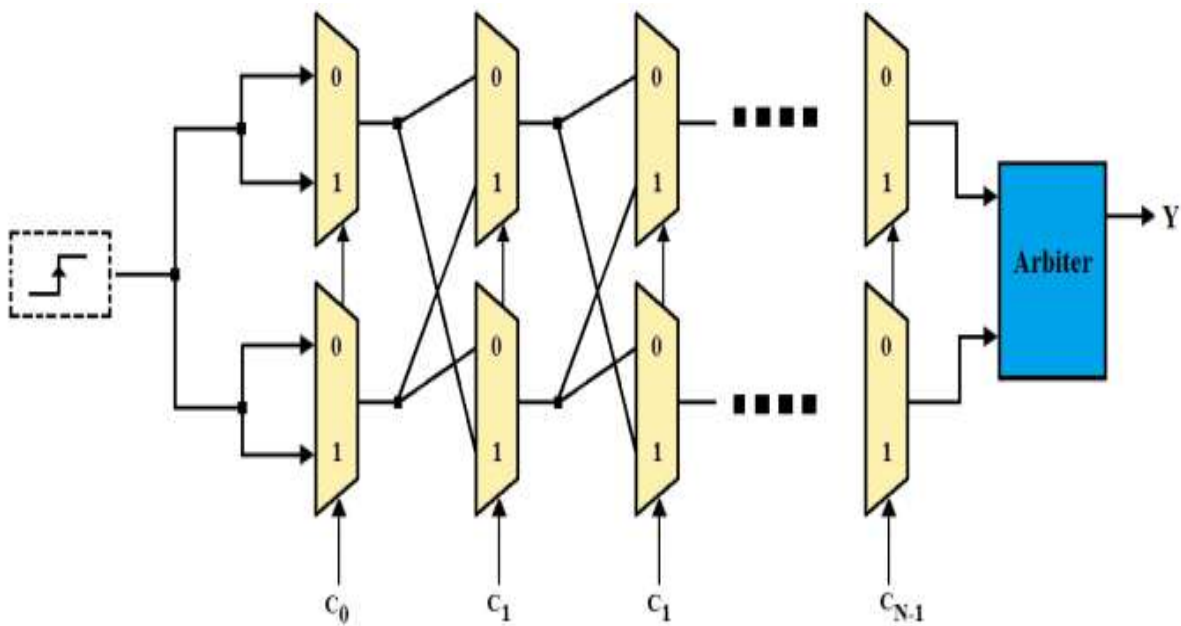

**Fig 1. Design of Arbiter PUF**

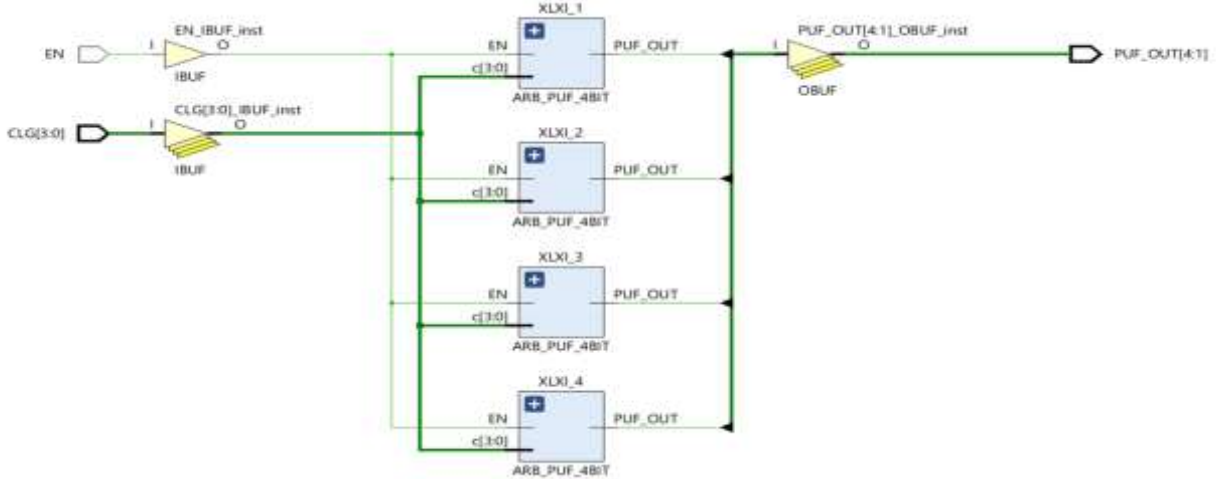

**Fig.2. Schematic of 4-bit Proposed Design**

## IV. ATTACK MODELLING

Machine learning (ML) based algorithms are a natural and powerful tool for modeling attacks. To compromise the PUF designs which are strong, it has the most promising attacking methods. In this technique, the collected data set of challenge-response pairs(CRPs) for a particular designed PUF is dived into two sets, one is used for train the model whereas the other is used to test the model to regenerate the designed PUF with a prediction of the response data set [2],[5].

Here, implemented one of that strong machine learning algorithms, logistic regression which helps to reproduce the PUF models with the collected data. The obtained results are discussed in the results section.

Logistic regression(LR) is one of the powerful ML-based algorithm and classification tools, it is a predictive analysis algorithm based on the concept of probability and it follows a sigmoid which lies in between 0 &1 as shown in fig(3). In this technique, the variables which are independent are given as input values and it produces the dependent variable which is output value/predicted value.

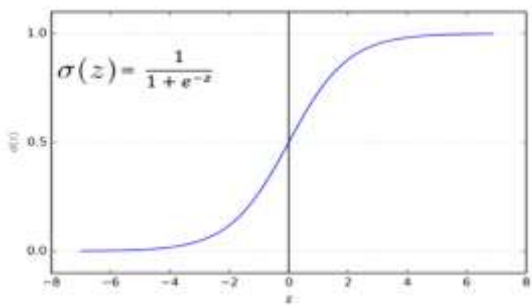

**Fig.3.Sigmoid Function**

## V. EXPERIMENTAL RESULTS

To get the desired results the design is implemented on the target device which is the FPGA board having a part number (xc7a35tcpg236-1). The design functionality was verified and records the different random sets of CRPs data [7] as shown in Table I. The tools used for implementation and processing are the Xilinx vivado design suite, and jupyter notebook .With the help of this recorded data, the LR model is designed and for different training rate sets generated prediction rates are recorded in Table II.

**Table I: Challenge-response pairs (CRPs) data set**

| CHALLENGES | RESPONSES |
|---|---|
| 64'h9283c630815977c | FF00FF0000FF00FF |
| 64'h824e3d711516856b | FFFF0000FFFFFF00 |
| 64'h92304516c4bb0240 | FF00FFFF0000FF00 |
| 64'h200fbac6d9bb7303 | 0000FFFFFF00FF00 |
| 64'h6844dcc6a582ac22 | 000000FFFFFFFFFF |
| 64'h686d3ec2141a7dfb | 00FFFFFF00FF0000 |
| 64'h6494978f8293cf35 | FF000000FF0000FF |
| 64'hef911feddf105f4e | 00FF00FF000000FF |
| 64'h7b2869d1d09564d2 | 0000FFFF00FFFFFF |
| 64'hf0d856b216b4c3a3 | FF00FFFFFF000000 |

Using the above-recorded data set, the model using logistic regression (LR) is designed, to obtain the prediction rate of the collected data, using a data split technique for different combinations of test and train and model is executed in the tool jupyter notebook. If the test rate is 0.25 for the data set, the remaining set is used for training and vice versa. With the help of the prediction rate (Pr) accuracy of the model is calculated. The ratio of the true predicted responses and test set which is predefined is named as the Prediction rate.

**Table II: Performance of logistic regression model with different test/train splits**

| TEST CRPs | 0.15 | 0.25 | 0.35 |
|---|---|---|---|
| | **Prediction Rate(Pr) %** | | |
| 750 | 55.75 | 53.19 | 53.24 |
| 1650 | 53.62 | 53.75 | 46.71 |
| 2850 | 50 | 51.05 | 51.5 |
| 4920 | 50.41 | 53.08 | 52.96 |

The average prediction rate was obtained by nearly around 52 percent. Regarding plot was shown in figure 4.

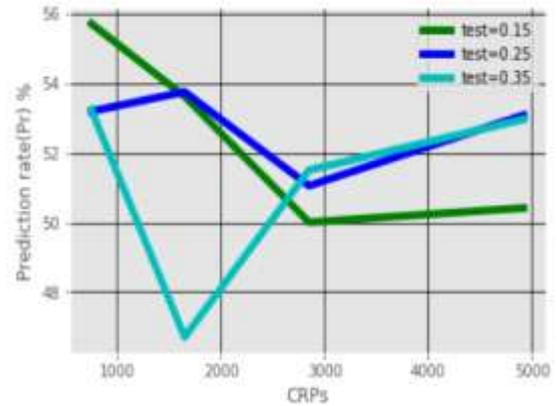

**Fig.4. Plot for the CRPs Vs Prediction rate**

The proposed design implementation and macro placement on the FPGA are shown in fig 5, 6 respectively.

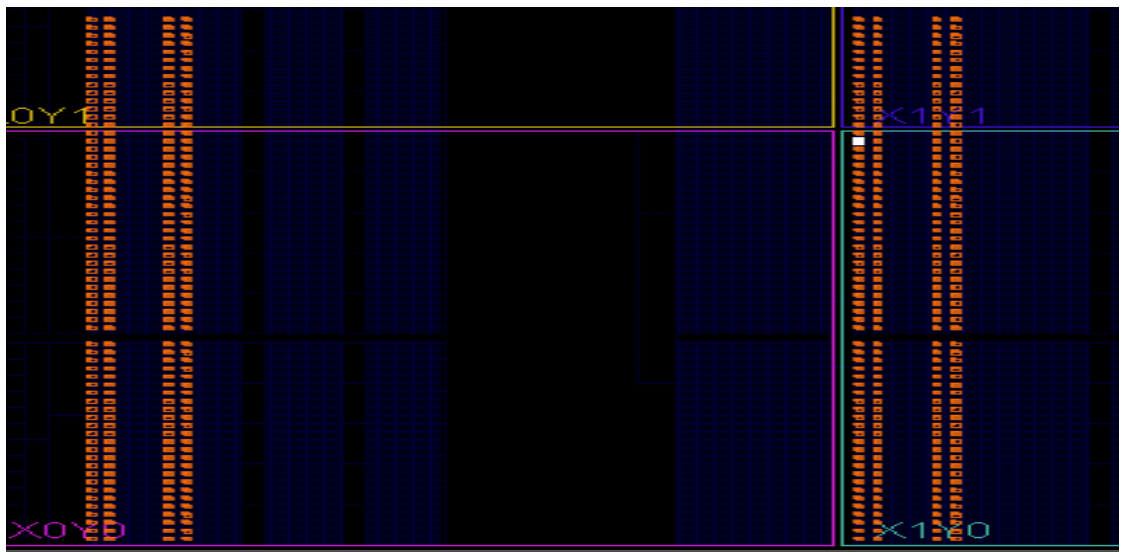

**Fig.5. Implementation of a 64-bit proposed design**

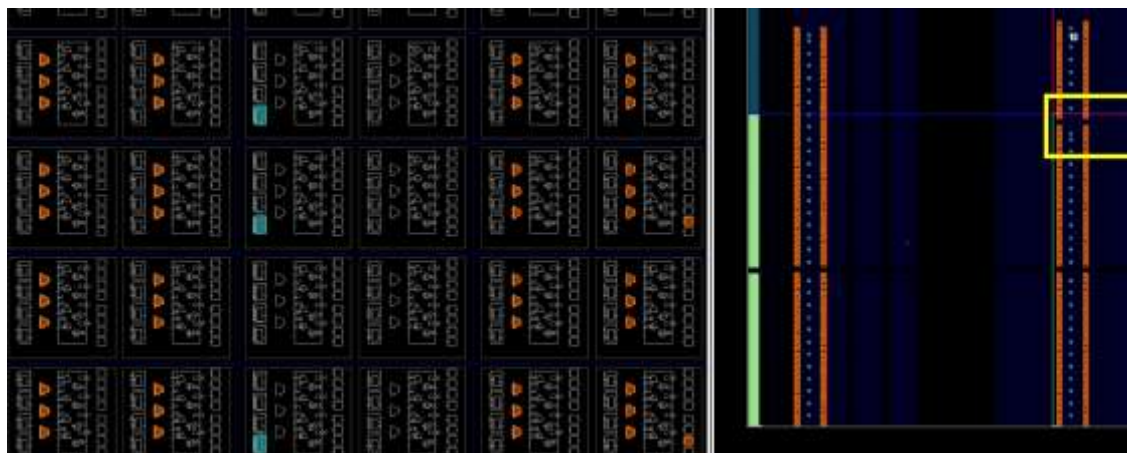

**Fig.6. 8-bit Macro placement on the FPGA fabric**

## VI. CONCLUSION

By modifying the arbiter based PUF design; a new design of PUF was proposed to improve the resistance capacity of design during attacks. For the proposed design different set of challenge-response pairs are obtained and comparatively got the good prediction rate with reduction of around 45 percent, on an average of around 52 percent is obtained, so it shows the design is more secure and strong enough to resist the attacks.

## ACKNOWLEDGMENT

I would like to acknowledge the support of my guide Mr.J.Dinesh Reddy, Assistant Professor, Department of E.C.E, BMSCE, for his guidance and encouragement throughout this work. This work is part of research work guided by Dr.K.P.Vittal, Professor, Department of E.E.E, NITK, and without his esteemed guidance and his support this wouldn't have taken its shape.

## REFERENCES

[1] J. Ye, Q. Guo, Y. Hu, H. Li, and X. Li, "*Modeling attacks on strong physical unclonable functions strengthened by random number and weak PUF*," in the 2018 IEEE 36th VLSI Test Symposium (VTS), San Francisco, CA, USA, 2018 pp. 1-6.

[2] R ührmair, Ulrich et al. "*Modeling attacks on physical unclonable functions.*" ACM Press, 2010. 237.

[3] D. P. Sahoo, R. S. Chakraborty, and D. Mukhopadhyay, "*Towards Ideal Arbiter PUF Design on Xilinx FPGA: A Practitioner's Perspective*," 2015 Euromicro Conference on Digital System Design, Funchal, 2015, pp. 559-562.

[4] U. Rührmair and D. E. Holcomb, "*PUFs at a glance*," 2014 Design, Automation & Test in Europe Conference & Exhibition (DATE), Dresden, 2014, pp. 1-6.

[5] S. Kumar and M. Niamat, "*Machine Learning-based Modeling Attacks on a Configurable PUF*," NAECON 2018 - IEEE National Aerospace and Electronics Conference, Dayton, OH, 2018, pp. 169-173.

[6] J. Ye, Y. Hu and X. Li, "*RPUF: Physical Unclonable Function with Randomized Challenge to resist modeling attack*," 2016 IEEE Asian Hardware-Oriented Security and Trust (Asian HOST), Yilan, 2016, pp. 1-6.

[7] Q. Guo, J. Ye, Y. Gong, Y. Hu and X. Li, "*Efficient Attack on Non-linear Current Mirror PUF with Genetic Algorithm*," 2016 IEEE 25th Asian Test Symposium (ATS), Hiroshima, 2016, pp. 49-54.

[8] A. Roelke and M. R. Stan, "*Attacking an SRAM-Based PUF through Wear out*," 2016 IEEE Computer Society Annual Symposium on VLSI (ISVLSI), Pittsburgh, PA, 2016, pp. 206-211.

[9] Mehboob Hasan Ahmed, Rutuja Jagtap, Roopal Pantode and Prof. S. S. Phule, *"An FPGA Chip Identification Generator using Configurable Ring Oscillator"* SSRG International Journal of Electronics and Communication Engineering 3.4 (2016): 10-14.